\documentclass[twocolumn,aps]{revtex4}

\newcommand {\bea}{\begin{eqnarray}}
\newcommand {\eea}{\end{eqnarray}}
\newcommand {\be}{\begin{equation}}
\newcommand {\ee}{\end{equation}}

\usepackage{graphicx}

\begin{document}


\title{The Kohn-Luttinger Effect in Gauge Theories}

\author{T.~Sch\"afer}

\affiliation{Department of Physics, North Carolina State University,
Raleigh, NC 27695}

\begin{abstract}
Kohn and Luttinger showed that a many body system of fermions
interacting via short range forces becomes superfluid even if 
the interaction is repulsive in all partial waves. In gauge 
theories such as QCD the interaction between fermions is long 
range and the assumptions of Kohn and Luttinger are not satisfied. 
We show that in a $U(1)$ gauge theory the Kohn-Luttinger phenomenon 
does not take place. In QCD attractive channels always exist, but 
there are cases in which the primary pairing channel leaves some
fermions ungapped. As an example we consider the unpaired fermion 
in the 2SC phase of QCD with two flavors. We show that it acquires 
a very small gap via a mechanism analogous to the Kohn-Luttinger effect. 
The gap is too small to be phenomenologically relevant. 
\end{abstract}

\maketitle

\section{Introduction}
\label{sec_intro}

 The conventional picture of superconductivity in cold Fermi 
liquids is that attractive interactions between fermions lead to 
a pairing instability that destroys the Fermi surface. Bardeen, 
Cooper, and Schrieffer showed that this instability is present 
for any attractive interaction, no matter how weak \cite{BCS:1957}.
This would seem to imply that a Fermi liquid with purely 
repulsive interactions is stable and has a Fermi surface. 

 Kohn and Luttinger \cite{Kohn:1965} showed that in the case 
of fermions interacting via short range forces this is not correct. 
There is a pairing instability even if the interaction is repulsive 
in all partial wave channels. The resulting critical temperature is 
typically very small, but the Kohn-Luttinger effect may well play a 
role in $p$-wave pairing in liquid $^3$He \cite{Baranov:1992,Efremov:2000}
and cold atomic gases \cite{Bulgac:2006gh} as well as $d$-wave pairing 
in high $T_c$ superconductivity.

 In this paper we study whether the Kohn-Luttinger effect also 
takes place in gauge theories. This study is part of a broader 
program to understand to what extent results from Fermi liquid 
theory can be extended to dense systems of fermions interacting 
via long range gauge forces. Non-Fermi liquid effects in the 
specific heat and the fermion dispersion relation were studied
in \cite{Holstein:1973,Baym:1975va,Polchinski:ii,Chakravarty:1995,Ipp:2003cj,Schafer:2004zf}.
The specific question whether repulsive interactions in gauge theories
can lead to superfluidity is of interest in connection with effective
$U(1)$ gauge theories that arise in condensed matter systems. In 
non-abelian gauge theories attractive channels always exist. However, 
there are a number of superconducting phases in which some of the 
fermions remain ungapped and the residual interactions between these 
excitations are repulsive. A simple example is color superconductivity
in QCD with two flavors. In this case up and down quarks with two
different colors form an $s$-wave superconductor whereas up and 
down quarks of the third color remain gapless. Gapless fermions
also appear in the gapless color-flavor-locked (CFL) phase
\cite{Alford:2003fq}.
 
\section{Short range forces}
\label{sec_short}

 Kohn and Luttinger considered short range interactions characterized
by a BCS potential $V(x)$. This potential describes particle-particle
scattering $\vec{p}_1+(-\vec{p}_1)\to \vec{p}_3+(-\vec{p}_3)$ with 
$|\vec{p}_1|=|\vec{p}_3|=p_F$ and $x=\hat{p}_1\cdot\hat{p}_3$ is the
scattering angle. The partial wave amplitudes are
\be
V_l = \int dx\ P_l(x)V(x),
\ee
where $P_l(x)$ is the Legendre polynomials of order $l$. If one of the 
partial wave amplitudes is attractive then $s$-channel rescattering 
between the particles will lead to a logarithmic growth of the interaction 
as the energy of the pair goes to zero. The system becomes superfluid when 
the effective interaction reaches a Landau pole. If all $V_l$ are repulsive 
then the effective interaction goes to zero logarithmically near the Fermi 
surface.

\begin{figure}
\includegraphics[width=8cm,clip=true]{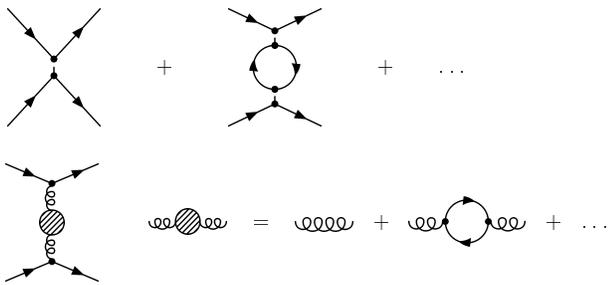}
\caption{\label{fig_resc}
The upper panel shows the direct interaction and the $t$-channel 
particle-hole correction in a theory with short range interactions. 
The lower panel shows the direct interaction in a gauge theory. In
this case, $t$-channel particle-hole bubbles have to be resummed
at leading order.  }
\end{figure}

 Kohn and Luttinger observed that in this case $t$-channel particle-hole
exchanges cannot be neglected. They note that if the BCS amplitude
and all its derivatives are analytic then the tree level partial wave 
amplitudes have to be exponentially small, $V_l\sim e^{-l}$, as $l\to
\infty$. This implies that perturbative corrections to the effective 
interaction can be important even if the coupling is small. In particular, 
if perturbative corrections are non-analytic in the scattering angle $x$
then they will dominate over the tree level amplitude for sufficiently
large $l$. The $t$-channel particle-hole bubble shown in 
Fig.~\ref{fig_resc} is
\be 
\Pi(\nu\!=\!0,q) = \frac{mk_F}{4\pi^2}
  \left\{ -1 +\frac{1}{q}\bigg(1-\frac{q^2}{4}\bigg) \log\bigg(
    \frac{2-q}{2+q} \bigg) \right\},
\ee
where $\nu=k_0$ is the energy transfer and $q=|\vec{k}|/k_F$ is the 
momentum transfer in units of $k_F$. This expression has a logarithmic
singularity at $q=2$ which corresponds to $x=-1$. The contribution 
of the singularity to the BCS amplitude is 
\be 
\delta V(x) = -\left[V(-1)\right]^2 \frac{mk_F}{16\pi^2}
   (1+x) \log(1+x) .
\ee
The correction to the partial wave amplitudes can be determined by 
using the generating functional 
\be 
P_l(x) = \frac{1}{2^l l!}\frac{d^l}{dx^l} \left[
  \left(x^2-1\right)^l\right]
\ee
and integrating by parts $l$ times. We find
\be 
\delta V_l =  (-1)^{l+1}\frac{mk_F}{4\pi^2} \frac{[V(-1)]^2}{l^4}
\ee
which is attractive for even $l$ and falls off as a power of $l$. This 
implies that even if all $V_l$ are repulsive there is a critical $l_{crit}$ 
such that $V_l+\delta V_l$ is attractive for $l$ even and $l>l_{crit}$. 

 In nuclear physics the screening correction is known as the 
``induced interaction''. Consider a short rage potential with 
a strength adjusted to give the the scattering length $a$. If $a$ is 
negative then the $s$-wave interaction is attractive and in BCS theory 
we find a gap 
\be 
 \Delta = \frac{8E_F}{e^2}\exp\left(-\frac{\pi}{2k_F|a|}\right),
\ee
where $k_F$ is the Fermi momentum and $E_F$ is the Fermi energy. 
The screening correction reduces the magnitude of the gap by a 
factor $(4e)^{1/2}\simeq 2.2$ \cite{Gorkov:1961}. If the $a$ is 
positive then the interaction is repulsive and no $s$-wave pairing 
takes place. However, the interaction is attractive in higher 
partial waves. The $d$-wave gap is of order
\be 
 \Delta \sim E_F \exp\left(-\frac{\pi^2}{(2k_Fa)^2}
      \frac{105}{2(8-\log(2048))} \right)
\ee
Higher order corrections to this result were studied in 
\cite{Baranov:1992,Efremov:2000}.

\section{Superconductivity in gauge theories}
\label{sec_csc}

 In gauge theories the interaction is long range and the 
particle-hole screening correction has to be included at leading 
order even if the angular momentum $l$ is small. The screened electric 
and magnetic gauge boson propagators are given by 
\cite{Kapusta:1989,LeBellac:1996}
\bea 
iD_{00}(\omega,\vec{k}) &=& \frac{1}{\vec{k}^2+2m^2}, \\
iD_{ij}(\omega,\vec{k}) &=& \frac{\delta_{ij}-\hat{k}_i\hat{k}_j}
 {\vec{k}^2-i\frac{\pi}{2}m^2 \omega/|\vec{k}|},
\eea
where  $m^2$ is the dynamically generated gauge boson mass and we have 
assumed that $|\omega|\ll|\vec{k}|$. In perturbation theory $m^2=
g^2N_f\mu^2/(4\pi^2)$ where $g$ is the gauge coupling, $\mu$ is the 
baryon chemical potential and $N_f$ is the number of fermions in the 
fundamental representation. The electric interaction is completely 
screened. At low energy multipole moments of the interaction are 
independent of energy and decrease exponentially with $l$. The magnetic 
interaction, on the other hand, in only dynamically screened. Multipole
moments diverge as a logarithm of energy and the coefficient of 
$\log(\omega)$ is independent of $l$. We find \cite{Son:1998uk}
\be 
\label{fl_om}
 f_l(\omega) =-c_A \frac{2g^2}{3} \log(\omega) ,
\ee
where $c_A$ is a symmetry factor that depends on the channel and
the gauge group. For color anti-symmetric quark pairing in QCD 
$c_A=-(N_c+1)/(2N_c)$. The renormalization group flow of the interaction
equ.~(\ref{fl_om}) was first studied by Son \cite{Son:1998uk}. The value 
of the $s$-wave gap is 
\be
\label{gap_oge}
\Delta_0 \simeq 2\Lambda_{BCS}
   \exp\left(-\frac{\pi^2+4}{8}\right)
   \exp\left(-\frac{3\pi^2}{\sqrt{2}g}\right).
\ee
where the scale is given by $\Lambda_{BCS}=256\pi^4(2/N_f)^{5/2}g^{-5}\mu$
and we have set $N_c=3$. The coefficient of $1/g$ in the exponent is 
determined by equ.~(\ref{fl_om}) and is independent of $l$. The 
pre-exponential factor is sensitive to electric gauge boson exchanges and
fermion self-energy corrections and depends on $l$. The dependence 
on $l$ can be determined by expanding the magnetic and electric gauge
boson exchange amplitudes around the forward direction. The result is 
$\Delta_l=\exp(3c_l)\Delta_0$ with \cite{Schafer:2000tw,Brown:1999yd}
\be 
 c_l= \int \frac{dx}{1-x}\ \left( P_l(x) -1 \right).
\ee
The coefficients $c_l$ can be computed using the recursion relation 
for the Legendre polynomials. We find 
\be 
 c_l = -2 \sum_{n=1}^l \frac{1}{n} 
 \;  \stackrel{l\to\infty}{\longrightarrow} \; -2\log(l) 
\ee
which implies that the gap scales as $\Delta_l\sim \Delta_0/l^6$ 
and becomes very small for $l\gg 1$. We should note, however, that 
the result is reliable only if $\log(l)< 1/g$. When this inequality 
is violated we cannot expand the scattering amplitude around $x=1$. 
In practice, this is already the case for $l\sim 2$. 

\begin{figure}
\includegraphics[width=7cm,clip=true]{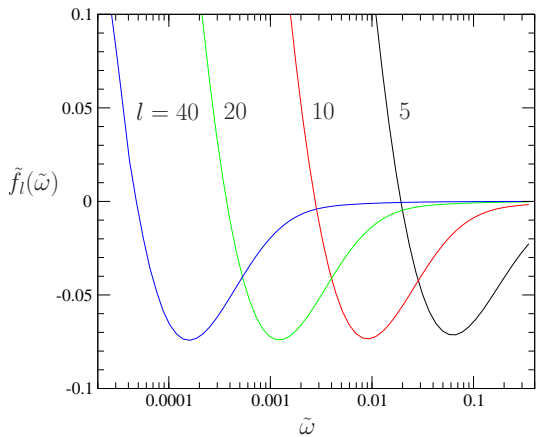}
\caption{\label{fig_fl}
Scaled gauge boson exchange multipole amplitudes $\tilde{f}_l(\tilde\omega)$ 
as a function of the dimensionless energy variable $\tilde\omega$. The 
figure shows the $l=5,10,20,40$ multipoles. For QCD at $\mu=500$ MeV 
the energy scale corresponds to roughly 1 GeV. }
\end{figure}

 This suggests that the behavior of the scattering amplitude changes
when $\log(l)\sim 1/g$, and that something analogous to the Kohn-Luttinger
effect may take place. In order to study this question we compute the 
partial wave amplitudes without relying on a expansion around $x=1$.
The transverse gauge boson exchange contribution is given by
\be 
 f_l(\omega) = c_R g^2 \tilde{f}_l(\tilde\omega), \hspace{1cm}
\tilde\omega = \frac{\pi}{4\sqrt{2}}\frac{m^2}{\mu^2}
  \left(\frac{\omega}{\mu}\right)
\ee
where $c_R$ is a color factor and
\be 
 \tilde{f}_l(\tilde\omega)  = \int dx\ 
  \frac{P_l(x)}{1-x +\tilde\omega/\sqrt{1-x}} .
\ee
In Fig.~\ref{fig_fl} we show the function $\tilde{f}_l(\tilde\omega)$ 
for several different values of $l$. We observe that for every value
of $\tilde\omega$ the scattering amplitude changes sign if $l$ is 
larger than some critical value. Also, for every value of $l$ ($l\geq 2$)
the amplitude changes sign for $\tilde\omega>\tilde\omega_0(l)$. The
function $\tilde{f}_l(\tilde\omega)$ exhibits a very simple scaling 
behavior $\tilde{f}_l(\tilde\omega)=F(\tilde\omega_0(l)/\tilde\omega)$
with $\tilde\omega_0(l)\sim l^{-\alpha}$ and $\alpha\simeq 2/\log(2)$.

 The figure shows that even if the interaction in the forward direction
is repulsive higher multipole amplitudes are attractive for almost all 
energies. The question is whether this is sufficient in order to produce 
superconductivity. In the appendix we discuss conditions under 
which interactions that are repulsive in some energy regime 
induce pairing instabilities. We show that if the interaction 
is repulsive below some energy $\omega_0$ then a necessary condition
for pairing is that the gap $\Delta$ is larger than $\omega_0$.
An upper bound on the gap can be obtained by assuming that $\Delta
(\omega)$ is independent of $\omega$. In this case we find
\be 
\label{eli_rep}
 1 =-\frac{g^2c_R}{8\pi^2}\int_\Delta \frac{d\omega}{\omega}
   \tilde{f}_l(\omega)
\ee
and consistency requires that $\Delta > \omega_0(l)$. This means
that the integral on the RHS can be bounded by replacing the lower 
limit by $\omega_0(l)$. This integral is only weakly dependent on 
$l$ and quickly approaches a finite limiting value, see Fig.~\ref{fig_fl}. 
Numerically we find 
\be 
\label{fcrit}
\lim_{l\to\infty}\int_{\omega_0(l)} \frac{d\omega}{\omega}
   \tilde{f}_l(\omega) = -0.107 . 
\ee
This implies that in the limit of large $l$ a gap can only appear
if $0.107g^2c_R/(8\pi^2)>1$. We can ask if this result is modified 
by higher order perturbative corrections. If $\omega_0(l)$ is smaller 
than the scale of non-Fermi liquid effects in the normal phase, 
$\omega_{\it nfl}=m\exp(-9\pi^2/g^2)$, then self energy corrections to 
the fermion propagator have to be taken into account. These corrections 
further reduce equ.~(\ref{fcrit}). Higher order contributions to the 
vertex scale as $\omega^{1/3}$ and can be neglected \cite{Schafer:2004zf}.
We conclude that in the weak coupling limit superconductivity does 
not take place in repulsive channels. Indeed, judging from the weak
coupling result, superconductivity is not likely even for large 
values of the coupling. 

\section{Conclusions}
\label{sec_sum}

  Our results are directly applicable to a $U(1)$ gauge theory 
at finite density. In that case there are no attractive channels
and our results imply that superconductivity does not take place. 
Of course, if the density is very small or the coupling is very 
large then the system may exhibit other instabilities, for example
towards ferromagnetism or crystalline order \cite{Bloch:1929}. 

 In QCD attractive channels always exist. However, there are cases
in which the primary pairing channel leaves some modes ungapped.
The simplest example is the 2SC phase of $N_f=2$ QCD. In this 
case up and down quarks with two different colors are paired 
but quarks of the third color are unpaired. At energies below 
the primary pairing gap $\Delta$ these quarks interact via the 
exchange of the hypercharge gluon of the $SU(3)$ gauge group. 
This gluon is part of the broken sector of the gauge group, and
as a consequence the magnetic interaction is screened. The 
momentum dependent screening mass was computed in 
\cite{Rischke:2000qz,Rischke:2002rz}. The result is 
\be 
{\rm Re}\,\Pi_T(k_0,k) = \frac{\pi^2}{4}\, m^2\, \frac{\Delta}{k}
\hspace{0.5cm}(k_0\ll\Delta, \, k\gg\Delta).
\ee
For $k<\Delta$ the screening mass goes to a constant. We observe
that this result has the same structure as the Landau damping
term except that the self energy is real rather than imaginary 
and the energy is replaced by the gap. As a consequence the partial 
wave amplitudes are identical to the results shown in Fig.~\ref{fig_fl} 
for $\omega>\Delta$ and become constant for $\omega<\Delta$. This means 
that pairing takes place in the partial wave for which $\omega_0\sim 
\Delta$, so $l\sim (m/\Delta)^{1/\alpha}$. This is parametrically large, 
but for values of the gap that are of physical interest $l$ could
be as small as $l=2$. 

 Finally we can try to estimate the magnitude of the pairing
gap. The scale for the secondary gap $\Delta_R$ is set by the 
primary gap $\Delta$. The multipole amplitude $f_l(\omega)$ is 
independent of energy for $\omega<\Delta$ and the dependence of 
the gap on the coupling constant is given by the BCS result,
$\Delta_R\sim \Delta\exp(-c/g^2)$. The constant $c=8\pi^2/(c_R
|f_l(\Delta)|)$ is determined by the color factor $c_R$ and the 
partial wave amplitude at the scale set by the primary gap. 
Using $c_R=1/3$ for the ungapped quark in the 2SC phase and
$|f_l(\Delta)|\leq 0.07$ from Fig.~\ref{fig_fl} we get $c\simeq
3300$. This means that that the gap is completely negligible, even 
if the coupling is strong. In practice the most likely mechanism 
for generating a secondary gap is the attractive instanton interaction 
in the spin-one channel $\langle uC\vec{\alpha}d\rangle$ discussed
in \cite{Alford:1997zt}. This mechanism requires strong coupling, 
since perturbative interactions in this channel are repulsive.

Acknowledgments: I would like to thank I.~Shovkovy for a useful 
discussion. This work was supported by US Department of Energy 
grants DE-FG02-03ER41260.

\appendix
\section{Energy dependent interactions}
\label{sec_app}

 In this appendix we study how superconductivity can arise in 
microscopic models that contain an effective interaction which 
depends on energy and is repulsive near the Fermi surface. Consider 
a gap equation of the type 
\be 
\Delta(\omega) = -\int d\nu \ \frac{V(\omega,\nu)\Delta(\nu)}
   {\sqrt{\nu^2+\Delta(\nu)^2}} .
\ee
Clearly, this equation has a non-trivial solution if the potential 
is attractive at low energy. However, non-trivial solutions can also 
arise if the potential is repulsive. Consider a potential that is 
repulsive everywhere, but less repulsive for $\omega<\omega_D$ \cite{Nayak},
\be 
 V(\omega,\nu) = \left\{ \begin{array}{cl}
  \; V_R-V_A \; & |\omega-\nu| < \omega_D, \\
         V_R    & |\omega-\nu| > \omega_D.
  \end{array}\right.
\ee
A natural ansatz for the gap is 
\be 
 \Delta(\omega) = \left\{ \begin{array}{cl}
   \; \Delta_1 \; & |\omega| < \omega_D \\
   \; \Delta_2 \; & |\omega| > \omega_D
 \end{array}\right. 
\ee
which leads to two coupled equations 
\bea
 \Delta_1 &=& -(V_R-V_A)\int_0^{\omega_D} d\omega
   \frac{\Delta_1}{\sqrt{\omega^2+\Delta_1^2}} \nonumber \\
 & & \hspace{1.5cm}\mbox{} 
   - V_R \int_{\omega_D}^\Lambda d\omega
   \frac{\Delta_2}{\sqrt{\omega^2+\Delta_2^2}} \\
 \Delta_2 &=& \; -V_R \; \int_0^{\omega_D} d\omega
   \frac{\Delta_1}{\sqrt{\omega^2+\Delta_1^2}} \nonumber \\
 & & \hspace{1.5cm}\mbox{} 
 - V_R \int_{\omega_D}^\Lambda d\omega
   \frac{\Delta_2}{\sqrt{\omega^2+\Delta_2^2}} 
\eea
This system of equations has a non-trivial solution if the effective 
interaction $V_{\it eff}$ 
\be 
 V_{\it eff} = V_A - \frac{V_R}{1+V_R\log(\Lambda/\omega_D)}
\ee
is positive. This result has a simple interpretation from the 
point of view of the renormalization group. The repulsive interaction
is screened as the energy is lowered from the cutoff scale $\Lambda$
to the low energy scale $\omega_D$. If $V_A$ is larger than the 
screened $V_R$ then $V_{\it eff}$ is attractive. The gaps are $\Delta_1 
= 2\omega_D\exp(-1/V_{\it eff})$ and
\be
\Delta_2 = -\Delta_1\, 
    \frac{V_R}{V_{\it eff}(1+V_R\log(\Lambda/\omega_D))}.
\ee
The other situation of interest is the case of a potential that is
repulsive near the Fermi surface, but attractive at higher energies
\be 
 V(\omega,\nu) = \left\{ \begin{array}{cl}
  \; V_R \; & |\omega-\nu| < \omega_D , \\
    -V_A    & |\omega-\nu| > \omega_D .
  \end{array}\right. 
\ee
In this case the scale is set by the gap $\Delta$ for the purely 
attractive case $V_R=-V_A$. If $\Delta<\omega_D$ then there is no 
non-trivial solution for $V_R\geq 0$. If, on the other hand, $\Delta>
\omega_D$ then there is a non-vanishing gap even if $V_R>0$ provided
that $V_R$ remains perturbative, $V_R<1$. Again this result is easy
to understand in the context of the renormalization group. If the 
RG reaches a Landau pole above $\omega=\omega_D$ then the repulsive
low energy potential does not modify the evolution.


\end{document}